\documentclass{llncs}

\usepackage{amssymb,lscape}
\usepackage{graphicx}
\usepackage{enumitem}
\usepackage{multirow}
\usepackage{hhline}
\usepackage{url}
\usepackage{bbding}

\urldef{\mailsa}\path|lucas.gren@lucasgren.com|
\urldef{\mailsb}\path||
\urldef{\mailsc}\path||
\newcommand{\keywords}[1]{\par\addvspace\baselineskip
\noindent\keywordname\enspace\ignorespaces#1}

\begin{document}

\mainmatter  
\title{Understanding Work Practices of Autonomous Agile Teams: A Social-Psychological Review}
\titlerunning{Understanding Work Practices of Autonomous Agile Teams}

\author{Lucas Gren}
\authorrunning{Lucas Gren}
\institute{Chalmers $|$ University of Gothenburg, Gothenburg, Sweden\\
and Volvo Cars, Gothenburg, Sweden\\
\mailsa\\
\mailsb\\
\mailsc\\}

\toctitle{Lecture Notes in Computer Science}
\tocauthor{Authors' Instructions}
\maketitle

\begin{abstract}
The purpose of this paper is to suggest additional aspects of social psychology that could help when making sense of autonomous agile teams. To make use of well-tested theories in social psychology and instead see how they replicated and differ in the autonomous agile team context would avoid reinventing the wheel. This was done, as an initial step, through looking at some very common agile practices and relate them to existing findings in social-psychological research. The two theories found that I argue could be more applied to the software engineering context are social identity theory and group socialization theory. The results show that literature provides social-psychological reasons for the popularity of some agile practices, but that scientific studies are needed to gather empirical evidence on these under-researched topics. Understanding deeper psychological theories could provide a better understanding of the psychological processes when building autonomous agile team, which could then lead to better predictability and intervention in relation to human factors.
\end{abstract}

\section{Introduction}\label{sec:introduction}
The importance of understanding team autonomy has increased in the last decades due to agile development processes \cite{moe2019trends}. There have been studies on the barriers of self-organization in agile teams \cite{moe2008understanding}, the emerging roles of self-organizing agile teams and how these roles enable agility \cite{hoda2012self}, and the role of senior management \cite{hoda2011supporting} to just mention a few. Some authors, like Moe et al.\ \cite{moe2009understanding}, suggest using theories from social psychology to better understand team autonomy. They refer to studies on self-organization in psychology but here are many more theories in that field that would make sense to use in the software development context. The two theories from social psychology, namely Social Identity Theory and Group Socialization Theory where selected from a textbook on social psychology \cite{hogg2014sp}. I do not consider these theories more important than others, they were only selected based on the vast number of studies on them in the last decades. Therefore, they seem to be quite robust and relevant to investigate in the software development context. Some results might replicate, but others might not.

In order to theoretically analyze these two theories from social psychology and their connection to agile teams, I use the study by So \cite{so2010making} in which the author divide common agile practices into \emph{core}, \emph{technical}, \emph{team interaction}, and \emph{customer interaction} practices. Due to saving space, I selected only the \emph{core} and \emph{team interaction} practices (five in total) for this paper. I do not consider these practices to cover all aspects of agility nor to be the most important agile practices, however, the practices chosen are widely used in industry \cite{7890614}. 

I will describe the general agile work practices and connect these to existing social, management, and organizational psychology findings, but I will start by presenting the two important psychological theories that are in focus.

\section{Important Psychological Theories} \label{psy}
I have chosen to focus on two popular theories in social psychology on which none or very few studies exist in software engineering research. The first one is group socialization, which can be defined as the ``dynamic relationship between the group and its members that describes the passage of members through a group in terms of commitment and of changing roles'' \cite{hogg2014sp}. The idea is that a new team members will go through a certain set of phases through the group's lifespan. The group as a whole will evaluate a new member first by assessing how much a potential new member can contribute to the group's goal-fulfillment. The individual will also assess how much the group can fulfill their personal needs. Step two is commitment, which takes the outcome from the evaluation as input and is an assessment of both parties' beliefs about the rewadingness (i.e.\ the quality of being rewarding) of the relationship (and other alternative ones). The last phase is role transition in which the commitment reaches a critical level and the relationship thereby changes. These three phases are continuously depending on the result of the assessment. The individual goes through five phases of group socialization: (1) Investigation, (2) Socialization, (3) Maintenance, (4) Resocialization, and (5) Remembrance. These phases have transition steps in-between that are: (1) Entry, (2) Acceptance, (3) Divergence, and (4) Exit \cite{levine1994group}.

Group socialization is separate from stages that the whole group goes through together. The most famous and used model of group development was introduced in 1965 by Tuckman \cite{tuckman1965developmental}. He integrated many theories and research findings into a model with the four stages Forming, Storming, Norming, and Performing. The forming stage is when the group is new and need to set the stage and figure our what it is supposed to do and who can do what. Storming is a conflict stage where people now feel safe to question the other team members, which is needed to figure out good group goals and good strategies of work. The Norming phase is when the group starts to set group norms of their collaboration and know how to organize to be productive, and the final stage, Performing, is when the team can focus the most of being productive because they have created a system of good collaboration and effective conflict resolution techniques. These phases are similar to the ones suggested by Agazarian and Gantt \cite{agazarian2003phases} in systems centered theory, and what Wheelan \cite{wheelandev} also suggested as an integrated model of group development.

Another well-researched approach to explaining many group phenomena is the social identity theory (see e.g.\ Hewstone et al.\ \cite{hewstone2002intergroup} or Hogg \cite{hogg2014sp}). Not only has the theory gained empirical evidence in social psychology research but also in quite recent research in social neuroscience (see e.g.\ van Bavel and Cunningham \cite{van2010social}). To understand that theory, we first need to understand the concepts on which it is based. Social categorization is the classification of people into different social groups, which is a deeply-rooted human trait, and a person's social identity is the part of the self that is derived from the various memberships we have in social groups. Social identity theory is, therefore, the theory of group membership and intergroup relations based on self-categorization, social comparison and a self-definition regarding in-group\footnote{A group that an individual is a member of \cite{hogg2014sp}} properties (i.e.\ a prototype\footnote{Cognitive representation of the typical\slash ideal defining features of a category \cite{hogg2014sp}}). Self-categorization is how we categorize ourselves and thereby construct a social identity \cite{hogg2000we}. According to the minimal group paradigm \cite{tajfel1971social}, even explicitly random group assignments trigger discriminatory behavior against an out-group\footnote{A group that an individual is not a member of \cite{hogg2014sp}}. The idea is that a successful intergroup bias creates or protects (high) in-group status, which provides a positive social identity (which in turn satisfies group-members' need for positive self-esteem\footnote{Feelings about and evaluations of oneself \cite{hogg2014sp}}). Researchers have successfully explained how groups gain positive self-esteem through intergroup bias but have been less successful when explaining intergroup bias motives due to threats or depressed self-esteem \cite{hewstone2002intergroup}. However, Hogg and Williams \cite{hogg2000we} suggest that competition for positive social identity characterizes intergroup behavior.

\section{Agile Practices and Social Psychology} \label{lr}
\subsection{Iterative Development -- A core practice of agile development}
Delivering in short iterations has high face validity, but when broken down, these ideas include a diversity of competences and dynamics needed by the agile team to deliver value in such short iterations. In more general management research, there has been more thorough research on which general work practices contribute to performance (see e.g.\ Combs et al.\ \cite{combs2006much}) and to successfully implement iterative development, the team must have a high degree and maturity of, for example, staffing, decentralized decision-making, and communication \cite{evans2005high}. So to understand the dynamics of iterative development, we should consider these confounding factors before we, as researchers, jump to conclusions about other found effects. 

\subsection{Iteration Planning -- A teamwork practice}
Obtaining empowered and motivated individuals that have the needed support to solve any given task together with high levels of trust, are all aspects known to be necessary \cite{mchugh2012agile} but are not always in place \cite{buchanan2008you}. Creating a shared vision has also been shown in research to be a key for success since the beginning of the 1990s and is one of the main components of transformational leadership \cite{bass1990transactional}. A shared vision is necessary since the team needs an overall goal to break down when planning the upcoming iteration. Regarding the importance of simplicity in agile is somewhat connected to the concept of reducing waste in lean manufacturing, together with the continued avoidance of doing unnecessary activities in the project (or process) life-cycle \cite{hicks2007lean}. To plan in such a way, the team must know the members' real competences and abilities, which also implies maturity in the development process and that the members of the group are committed and fully integrated into the group. With such prerequisites, understanding the group socialization process then becomes paramount when understanding how teams plan in short iterations.

\subsection{Stand-up meetings -- A teamwork practice}
Developers, but also business people and testers, should be on the same team and collaboratively work together through the whole project life-cycle (i.e.\ having cross-functional teams). When connecting the popularity of having cross-functional teams in the modern workplace (see e.g.\ Denison et al.\ \cite{denison1996chimneys}) to social identity theory, it becomes clear that it, in fact, decreases intergroup bias. Having these various organizational functions share their chores and issues often, would be expected to increase cohesion and understanding of the whole project through shared mental models, which have also gained initial empirical support \cite{stray2016daily}. Having social identity theory and intergroup bias as factors in software engineering research would then probably increase the explained variance.

\subsection{Retrospectives -- A teamwork practice}
The idea of a retrospective meeting is that the team should reflect on possible improvement points about their teamwork at the end of each iteration \cite{derby2006arm}. More generally, such reflective meetings are often called \emph{team debriefs}, and have been shown with scientific rigor to increase effectiveness \cite{tannenbaum2013team}. McHugh et al.\ \cite{mchugh2012agile} found that these types of meeting need work and careful guidance to function in their intended way also in software development. In a recent longitudinal study, Lehtinen et al.\ \cite{lehtinen2017recurring} showed that, initially, newly formed teams focus more on task progress and task outcome and, as the teams mature, they focus to a larger extent on process and cooperation. Such findings also relate the ``agility'' of a team to group socialization and group development since members of the group will behave differently depending on how well integrated they are in the team \cite{levine1994group}, meaning that a well-integrated individual will be more likely to perform retrospectives in the way they are intended. If the socialization process is not a part of understanding the dynamics of retrospective meetings, studies will have difficulty explaining and predicting patterns of behavior.

\subsection{Co-location  -- A teamwork practice} 
Having the team co-located in the same room with requirements as sticky notes on physical boards have been promoted by the agile community in order to, again, increase the velocity of the development in a rapidly changing environment. Many cases have been reported where the communication challenges of distributed teams have been satisfactory dealt with using modern technology and slightly different practices (see e.g.\ Berczuk \cite{berczuk2007back}). Another study showed that both agile and traditional projects have the same issues regarding co-location \cite{noll2010global}. All-in-all, every social aspect of building relationships will become more cumbersome with distance and implies that more effort is needed to mitigate these challenges \cite{alzoubi2016empirical}. Since the social problems are amplified with distance, failing to understand their influence in distributed agile teams will have even larger negative effects on teamwork. And since agile processes are dependent on the team as a working unit, understanding the social aspects of both distributed and co-located teams are a key to building effective agile teams.   


\section{Discussion and Implications} 
As we have seen in this review, there is a lot of overlap between existing knowledge of, and research on, the workplace in general and the agile practices. A few internal organizational examples being decreasing inter-group bias through cross-functional teams \cite{denison1996chimneys}, striving towards self-organization of teams in order to increase responsiveness to change \cite{evans2005high}, creating organizational citizenship behavior through shared visions \cite{bass1990transactional}, empowerment and trust \cite{wat2005equity}, and removing \emph{waste} in the process \cite{hicks2007lean}. All these aspect are of interest to agile software engineering researchers when trying to understand the development of software using agile teams because these theories might add explanatory power to the observed behaviors. 

However, the theory could be seen as complex and hard to grasp for people without any behavioral science education, which means researchers must first run experiments to gather empirical evidence in order to eventually build a theory of ``agility,'' and then provide scientifically founded and validated guidelines to practitioners. One large hurdle of achieving this, though, is the fact that an overwhelming majority of human factors research in software engineering is conducted by software engineering researchers interested in psychology and not psychology researchers interested in software engineering, which often means that the research findings have little depth and offer little new insights from a psychological perspective. I will not cite any studies here due to the fact that such studies were conducted with the best of intentions and do have high value in that they have highlighted the importance of looking at psychological factors in the software engineering domain, which was not the case at all before.  

Social identity theory could be utterly useful when navigating through the added complexity of the different social relationship surfacing in an agile project. Hogg and Williams \cite{hogg2000management} explicitly suggest a set of propositions for how social identity and self-categorization relate to the organizational context. One of their propositions is that changes in which out-groups the in-group compares itself to, will change the view of the group's own identity, including the properties of the ideal member (i.e.\ the prototype). Another proposition is that harmonious relations between different subgroups of the organization is best kept by recognizing both the subgroups (e.g.\ Quality Assurance Engineer, Software Developers, Software Tester, etc.) and other organizational constellations, including the teams and the company as a whole. This means that the cross-functional agile teams must recognize both the value of the team as a whole but also the different roles and make distinctions between them. All these aspects should be part of agile team measurements in the future in order to fully make sense of the agile team context. 

When looking at the descriptions of the agile practices overall, many of the internal practices seem to assume full group-membership seen from a group socialization perspective \cite{levine1994group}. They also assume the entire work-group to be mature from a developmental perspective \cite{tuckman1965developmental}. In order to fully understand the social-psychological components of the team-based workplace in general, and the agile context in particular, we also need to investigate the temporal perspective of the interplay between group development, group socialization, and the agile approach to projects by setting up autonomous teams.

As have also been shown in this short review, the prescribed behavior in these agile practices are well-founded in social psychology, which provides social-psychological reasons for their popularity. The reason for this is that if the agile practices enable mechanisms known to work well for people in other contexts, it is likely that they would also be appropriate in some variation in the agile context. One example is the decrease of intergroup bias by having cross-functional teams. Therefore, I argue for that these theories should be applied more to the study of autonomous agile teams. In the review by \cite{dingsoyr20121213}, they also call for more theory-based research since the current status of the field mostly comprises method-specific case studies, which is particularly the case in software engineering studies on human factors. In this present study, I have explained some social-psychological underpinnings in relation to five common agile practices, which contributes to founding agile practices in more general social psychology theories. An understanding of such underpinnings can help abstract the essentials of agile software development as opposed to other approaches, but also guide researchers in conducting experiments using theory from social psychology in the software development context. Many of the agile principles are far from new in relation to human knowledge of work-groups. However, what might be considered as having gotten a stronger acceptance is the implementation of being responsive to change. The reasons for not relating agile software development to any existing science outside of software engineering might be due to lack of research knowledge from practitioners, but it might also reflect the difficulty of interdisciplinary research.

\section{Conclusion} 
Without understanding the psychology of groups, agile maturity survey findings are hard to use in order to improve one's own practices. Relating agile practices to deeper psychological theories, like in this study, could instead provide a deeper understanding of the psychological processes of implementing autonomous agile teams.

\bibliographystyle{splncs}
\bibliography{refssocial}

\begin{thebibliography}{10}

\bibitem{moe2019trends}
Moe, N.B., Stray, V., Hoda, R.:
\newblock Trends and updated research agenda for autonomous agile teams: a
  summary of the second international workshop at xp2019.
\newblock In: International Conference on Agile Software Development, Springer
  (2019)  13--19

\bibitem{moe2008understanding}
Moe, N.B., Dings{\o}yr, T., Dyb{\aa}, T.:
\newblock Understanding self-organizing teams in agile software development.
\newblock In: 19th Australian Conference on Software Engineering (aswec 2008),
  IEEE (2008)  76--85

\bibitem{hoda2012self}
Hoda, R., Noble, J., Marshall, S.:
\newblock Self-organizing roles on agile software development teams.
\newblock IEEE Transactions on Software Engineering \textbf{39}(3) (2012)
  422--444

\bibitem{hoda2011supporting}
Hoda, R., Noble, J., Marshall, S.:
\newblock Supporting self-organizing agile teams.
\newblock In: International Conference on Agile Software Development, Springer
  (2011)  73--87

\bibitem{moe2009understanding}
Moe, N.B., Dingsyr, T., Kvangardsnes, O.:
\newblock Understanding shared leadership in agile development: A case study.
\newblock In: 2009 42nd Hawaii International Conference on System Sciences,
  IEEE (2009)  1--10

\bibitem{hogg2014sp}
Hogg, M.A., Vaughan, G.M.:
\newblock Social Psychology. 7 edn.
\newblock Pearson, Harlow, England (2014)

\bibitem{so2010making}
So, C.:
\newblock Making software teams effective: {H}ow agile practices lead to
  project success through teamwork mechanisms.
\newblock Peter Lang, Frankfurt am Main (2010)

\bibitem{7890614}
Licorish, S.A., Holvitie, J., Hyrynsalmi, S., Leppänen, V., Spínola, R.O.,
  Mendes, T.S., MacDonell, S.G., Buchan, J.:
\newblock Adoption and suitability of software development methods and
  practices.
\newblock In: 23rd Asia-Pacific Software Engineering Conference (APSEC).
  (December 2016)  369--372

\bibitem{levine1994group}
Levine, J.M., Moreland, R.L.:
\newblock Group socialization: {T}heory and research.
\newblock European review of social psychology \textbf{5}(1) (1994)  305--336

\bibitem{tuckman1965developmental}
Tuckman, B.W.:
\newblock Developmental sequence in small groups.
\newblock Psychological bulletin \textbf{63}(6) (1965)  384--399

\bibitem{agazarian2003phases}
Agazarian, Y., Gantt, S.:
\newblock Phases of group development: Systems-centered hypotheses and their
  implications for research and practice.
\newblock Group Dynamics: Theory, Research, and Practice \textbf{7}(3) (2003)
  238

\bibitem{wheelandev}
Wheelan, S.:
\newblock Group processes: {A} developmental perspective. 2 edn.
\newblock Allyn and Bacon, Boston (2005)

\bibitem{hewstone2002intergroup}
Hewstone, M., Rubin, M., Willis, H.:
\newblock Intergroup bias.
\newblock Annual review of psychology \textbf{53}(1) (2002)  575--604

\bibitem{van2010social}
van Bavel, J.J., Cunningham, W.A.:
\newblock A social neuroscience approach to self and social categorisation: {A}
  new look at an old issue.
\newblock European Review of Social Psychology \textbf{21}(1) (2010)  237--284

\bibitem{hogg2000we}
Hogg, M.A., Williams, K.D.:
\newblock From {I} to we: {S}ocial identity and the collective self.
\newblock Group Dynamics: Theory, Research, and Practice \textbf{4}(1) (2000)
  ~81

\bibitem{tajfel1971social}
Tajfel, H., Billig, M.G., Bundy, R.P., Flament, C.:
\newblock Social categorization and intergroup behaviour.
\newblock European journal of social psychology \textbf{1}(2) (1971)  149--178

\bibitem{combs2006much}
Combs, J., Liu, Y., Hall, A., Ketchen, D.:
\newblock How much do high-performance work practices matter? {A} meta-analysis
  of their effects on organizational performance.
\newblock Personnel psychology \textbf{59}(3) (2006)  501--528

\bibitem{evans2005high}
Evans, W.R., Davis, W.D.:
\newblock High-performance work systems and organizational performance: {T}he
  mediating role of internal social structure.
\newblock Journal of management \textbf{31}(5) (2005)  758--775

\bibitem{mchugh2012agile}
McHugh, O., Conboy, K., Lang, M.:
\newblock Agile practices: {T}he impact on trust in software project teams.
\newblock IEEE Software \textbf{29}(3) (2012)  71--76

\bibitem{buchanan2008you}
Buchanan, D.A.:
\newblock You stab my back, {I}'ll stab yours: {M}anagement experience and
  perceptions of organization political behaviour.
\newblock British Journal of Management \textbf{19}(1) (2008)  49--64

\bibitem{bass1990transactional}
Bass, B.M.:
\newblock From transactional to transformational leadership: {L}earning to
  share the vision.
\newblock Organizational dynamics \textbf{18}(3) (1990)  19--31

\bibitem{hicks2007lean}
Hicks, B.J.:
\newblock Lean information management: {U}nderstanding and eliminating waste.
\newblock International journal of information management \textbf{27}(4) (2007)
   233--249

\bibitem{denison1996chimneys}
Denison, D.R., Hart, S.L., Kahn, J.A.:
\newblock From chimneys to cross-functional teams: {D}eveloping and validating
  a diagnostic model.
\newblock Academy of management journal \textbf{39}(4) (1996)  1005--1023

\bibitem{stray2016daily}
Stray, V., Sj{\o}berg, D.I., Dyb{\aa}, T.:
\newblock The daily stand-up meeting: {A} grounded theory study.
\newblock Journal of Systems and Software \textbf{114} (2016)  101--124

\bibitem{derby2006arm}
Derby, E., Larsen, D.:
\newblock Agile retrospectives: {M}aking good teams great.
\newblock Pragmatic Bookshelf, Raleigh, NC (2006)

\bibitem{tannenbaum2013team}
Tannenbaum, S.I., Cerasoli, C.P.:
\newblock Do team and individual debriefs enhance performance? {A}
  meta-analysis.
\newblock Human factors \textbf{55}(1) (2013)  231--245

\bibitem{lehtinen2017recurring}
Lehtinen, T.O., Itkonen, J., Lassenius, C.:
\newblock Recurring opinions or productive improvements --- {W}hat agile teams
  actually discuss in retrospectives.
\newblock Empirical Software Engineering \textbf{22}(5) (2017)  2409--2452

\bibitem{berczuk2007back}
Berczuk, S.:
\newblock Back to basics: {T}he role of agile principles in success with a
  distributed scrum team.
\newblock In: Agile Conference (AGILE), 2007, IEEE (2007)  382--388

\bibitem{noll2010global}
Noll, J., Beecham, S., Richardson, I.:
\newblock Global software development and collaboration: {B}arriers and
  solutions.
\newblock ACM Inroads \textbf{1}(3) (2010)  66--78

\bibitem{alzoubi2016empirical}
Alzoubi, Y.I., Gill, A.Q., Al-Ani, A.:
\newblock Empirical studies of geographically distributed agile development
  communication challenges: A systematic review.
\newblock Information \& Management \textbf{53}(1) (2016)  22--37

\bibitem{wat2005equity}
Wat, D., Shaffer, M.A.:
\newblock Equity and relationship quality influences on organizational
  citizenship behaviors: {T}he mediating role of trust in the supervisor and
  empowerment.
\newblock Personnel review \textbf{34}(4) (2005)  406--422

\bibitem{hogg2000management}
Hogg, M.A., Terry, D.I.:
\newblock Social identity and self-categorization processes in organizational
  contexts.
\newblock Academy of management review \textbf{25}(1) (2000)  121--140

\bibitem{dingsoyr20121213}
Dings{\o}yr, T., Nerur, S., Balijepally, V., Moe, N.B.:
\newblock A decade of agile methodologies: {T}owards explaining agile software
  development.
\newblock The Journal of Systems and Software \textbf{85} (2012)  1213—--1221

\end{thebibliography}
\newpage

\end{document}